\documentclass[epj]{svjour}
\usepackage{graphics}
\usepackage{lipsum}% this package is included to get dummy paragraphs for sample purpose.
\usepackage{graphicx,pstricks,epsfig,rotating,amsmath,amssymb}

\newcommand{\bea}{\begin{eqnarray}}
\newcommand{\eea}{\end{eqnarray}}
\newcommand{\apgt} {\ {\raise-.5ex\hbox{$\buildrel>\over\sim$}}\ }

\begin{document}
\title{A mixing interpolation method to mimic pasta phases in compact star matter}
\titlerunning{Mixing interpolation method to mimic pasta phases}
%\title{A new approach to mixed phases in compact star matter}
%\subtitle{Do you have a subtitle?\\ If so, write it here}
\author{David Blaschke\inst{1,2,3} \and David Alvarez-Castillo\inst{1,4} 
	\authorrunning{Blaschke and Alvarez-Castillo}
% \thanks is optional - remove next line if not needed
%\thanks{\emph{Present address:} Insert the address here if needed}%
}                     % Do not remove
%
%\offprints{}          % Insert a name or remove this line
%
\institute{
Bogoliubov Laboratory of Theoretical Physics, JINR Dubna, Dubna, Russia
\and 
National Research Nuclear University (MEPhI), Moscow, Russia
\and
Institute of Theoretical Physics, University of Wroclaw, Wroclaw, Poland
\and
Institute of Nuclear Physics, Polish Academy of Sciences, Cracow, Poland
}
\date{Received: 21 January 2020 / Accepted: 15 February 2020}
% The correct dates will be entered by Springer
%
\abstract{
We present a new method to interpolate between two matter phases that allows for a description of mixed phases and can be used, e.g., for mimicking transitions between pasta structures occurring in the crust as well as in the inner core of compact stars. 
This interpolation method is based on assuming switch functions that are used to define a mixture of subphases while fulfilling constraints of thermodynamic stability. 
The width of the transition depends on a free parameter, the pressure increment relative to the critical pressure of a Maxwell construction.
As an example we present a trigonometric function ansatz for the switch function together with a pressure increment during the transition. 
We note that the resulting mixed phase equation of state bears similarities with the appearance of substitutional compounds in neutron star crusts and with the sequence of transitions between different pasta phases in the hadron-to-quark matter transition. 
We apply this method to the case of a hadron-to-quark matter transition and test the robustness of the compact star mass twin phenomenon against the appearance of pasta phases modelled in this way.
 % end of PACS codes
} %end of abstract
\maketitle
\section{Introduction}
\label{intro}
Recently, in the discussion of the hadron-to-quark matter transition in compact stars, the actual character of the transition has become a matter of debate. 
Alternatives range from a strong first-order phase transition  to a crossover transition. 
In the absence of ab-initio lattice QCD studies of the equation of state (EoS) at zero temperature and high baryon densities, there is no guidance for the development of effective field theoretical models from this side.
One expects that both, quark deconfinement and chiral symmetry restoration shall take place in compact star interiors, but it is not clear whether these transitions occur simultaneously (as in finite temperature lattice QCD simulations at vanishing baryon density \cite{Bazavov:2016uvm}) or sequentially according to the idea of a quarkyonic phase 
(see \cite{Andronic:2009gj} and references therein) or massive quark matter phase \cite{Schulz:1987qg,Castorina:2010gy}.
The extrapolation from known nuclear matter properties around the saturation density may not be sufficiently reliable, as the extrapolation from high-density quark matter or even perturbative QCD matter asymptotics down to the hadronization region is equally insecure. 
In \cite{Kurkela:2014vha} these two limits have been joined by multi-polytrope EoS under the constraints that the speed of sound does not violate the causality constraint $c_s<c$ and the maximum mass of compact star configurations for the EoS shall reach at least 2 M$_\odot$ \cite{Antoniadis:2013pzd}. 
This still leaves a rather broad corridor in the compact star domain of the pressure versus energy density diagram, similar to the one reported earlier by Hebeler et al.~\cite{Hebeler:2013nza}. 
This even allows a strong phase transition with a large jump in energy density at a pressure $\sim 100$ MeV/fm$^3$ which would result in a third family branch of compact stars and the corresponding phenomenon of mass-twin star, as has been shown in \cite{Alvarez-Castillo:2017qki}.
The admissible EoS domain has recently been slightly narrowed \cite{Annala:2017llu} when the results from the gravitational wave measurement of the compact star merger GW170817 of the LIGO-Virgo Collaboration \cite{TheLIGOScientific:2017qsa} are used, but it still allows to accommodate a strong first order phase transition that can produce twin stars
\cite{Benic:2014jia}, at high and low masses as shown, e.g., in Ref.~\cite{Paschalidis:2017qmb}.  

Because of the absence of a reliable microscopic description in the relevant domain of (energy) densities, an interpolation technique has been suggested in \cite{Masuda:2012ed} to model the hadron-to-quark matter transition under neutron star constraints as a crossover transition.
For a recent, detailed discussion see Ref.~\cite{Baym:2017whm}. 
A similar interpolation technique has then been developed in \cite{Blaschke:2013rma} to model an unknown chemical potential dependence of nonlocal chiral model parameters, like the effective coupling strength in the vector meson channel.
This technique was further developed to a twofold interpolation method in order to account for a medium dependence of both, the vector coupling at high densities and confining effects in the vicinity of the hadronization transition \cite{Alvarez-Castillo:2018pve}.

An interpolation technique between hadronic and quark matter equations of state for compact star applications in generalization of a Maxwell construction has been developed recently in \cite{Ayriyan:2017tvl,Abgaryan:2018gqp} and serves to mimic pasta phase effects in the mixed phase of the hadron-to-quark matter transition in compact stars 
\cite{Ayriyan:2017nby,Blaschke:2019tbh}.
Such studies become particularly important when there is a large density jump involved in the transition so that surface tension and Coulomb effects of separately charged sub-phases in the globally charge neutral system may become important.
While such a replacement interpolation method (RIM) using a polynomial ansatz, in its simplest version in parabolic form, seems to provide a decent qualitative description of a mixed phase, there may be details of such a mixed phase construction which could call for a slightly different construction. 

\begin{figure}[!bhtp]
	\begin{center}
		\includegraphics[width=0.5\textwidth]{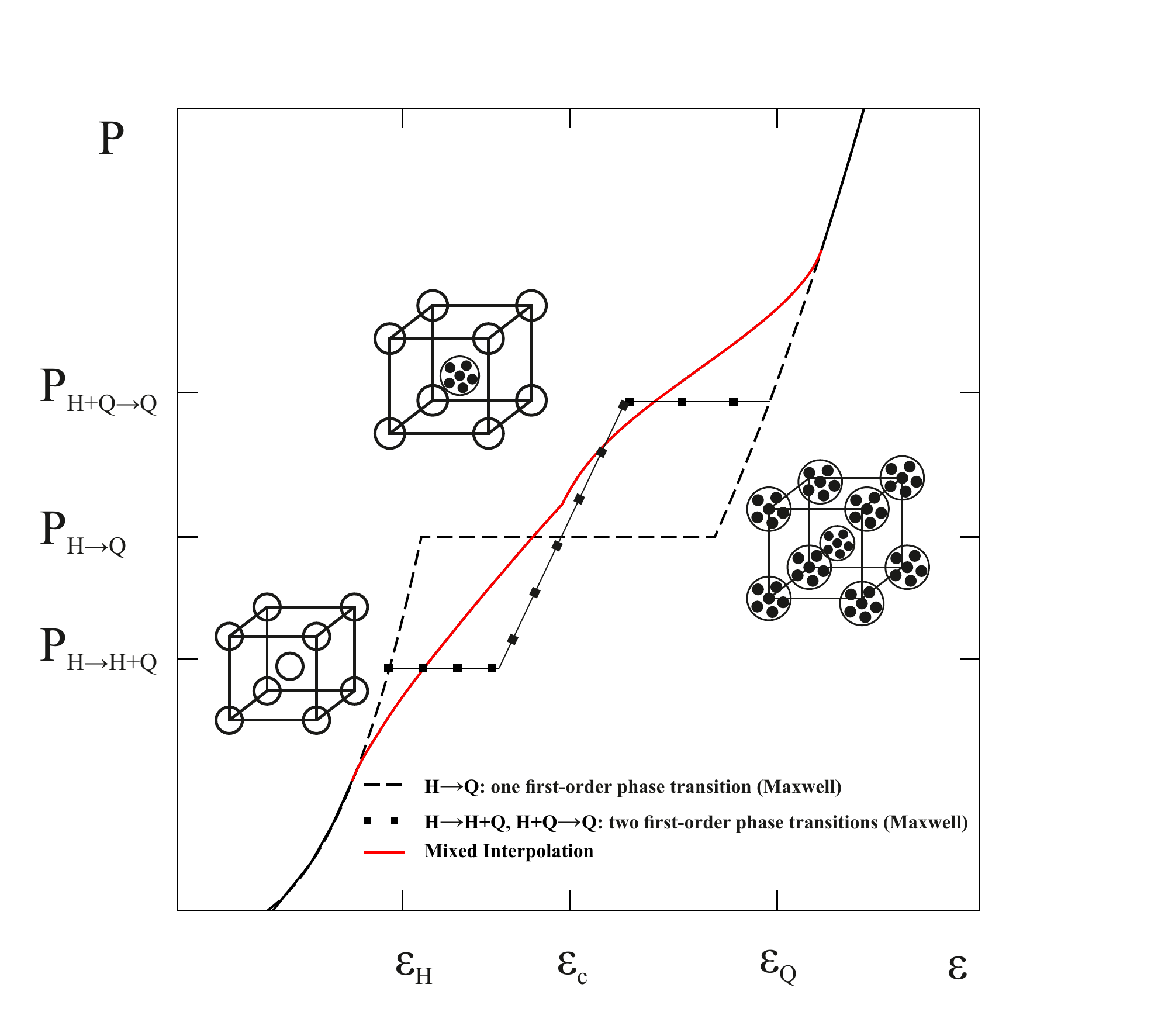}
	\end{center}
	%\vspace{1cm}
	\caption{\label{Compound} 
		 Illustration of the idea that a mixed phase with pasta structures may have similarities with substitutional compound structures in the neutron star crust, inspired by Fig.~1 of  Ref.~\cite{Chamel:2016ucp}. When hadronic and quark matter subphases are structured and form commensurate lattice structures in an intermediate density range, this "locking" of the structures may entail a transient stiffening of the EoS.} 
\end{figure}

Such an alternative construction we are suggesting in the present work. 
Its idea is not to invent an insertion of the pressure function between hadronic and quark matter asymptotes, but rather to use their functional forms in a mixing interpolation which in spirit is an averaging procedure with an additional pressure contribution from the structures in the mixed phase.
It is motivated by the construction of sequential transitions describing substitutional compounds in the neutron star crust \cite{Chamel:2016ucp}, see also Fig.~1 of Ref.~\cite{Blaschke:2018mqw}, which is adapted to the present case of the transition from hadron to quark matter via a structured mixed phase (pasta phase) in Fig.~\ref{Compound}.
The idea goes back to Freeman Dyson \cite{Dyson:1971} and was developed further in Ref.~\cite{Witten:1974}. 
As we shall describe in this work the mixing interpolation method (MIM) we suggest features a transient stiffening of the EoS similar to the case of the substitutional compounds and may in some cases provide a more accurate procedure to mimic the sequence of transitions between pasta structures in the mixed phase than the simple polynomial interpolation suggested in \cite{Ayriyan:2017tvl,Ayriyan:2017nby}.

\section{The mixing interpolation method (MIM)}
\label{sec:1}

\begin{figure}[!bhtp]
\begin{center}
\includegraphics[width=0.5\textwidth]{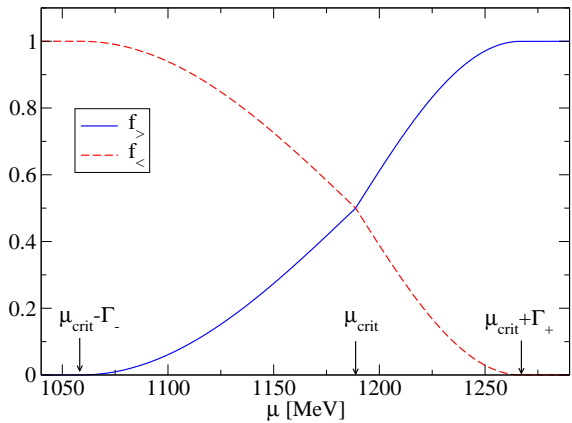}
\end{center}
%\vspace{1cm}
\caption{\label{Switches} Switch-on ($f_{>}$) and switch-off ($f_{<}$) functions that together with the pressure increment function $\Delta(\mu)$ are used to build an interpolation function that serves to describe a transition between two phases of matter that mimics the appearance of intermediate structures.} 
\end{figure}

\begin{figure}[!bhtp]
\begin{center}
\includegraphics[width=0.5\textwidth]{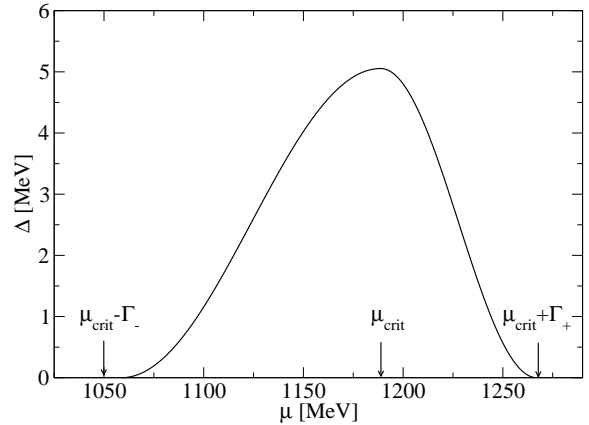}
\end{center}
\caption{\label{Delta}
Pressure increment function $\Delta(\mu)$ that together with the switch functions $f_{\lessgtr}(\mu)$ defines the mixing interpolation scheme of Eq.~(\ref{Interpolation}) describing the structured mixed phase between homogeneous phases.} 
\end{figure}

This method exploits the features of trigonometric functions, $\sin(x)$ and $\cos(x)$,
to interpolate smoothly between $1$ and $-1$ within a bounded domain of length $\pi$ with 
vanishing first derivative and normalized second derivative at these boundaries. 
The mixed interpolation between the hadronic (H) and quark (Q) EoS is introduced as
\begin{equation}
P(\mu)=P_{H}(\mu)f_{<}(\mu)+P_{Q}(\mu)f_{>}(\mu)+\Delta(\mu),
\label{Interpolation}
\end{equation}
where for the switch functions holds $f_{>}(\mu)=1-f_{<}(\mu)$. 
They are defined as
\[ 
f_{>}(\mu) =  \left\{
\begin{array}{lr}
      0, &  \mu<\mu_H\\
   \frac{1}{2}\left[1-\sin(\frac{\pi}{2}   \frac{\mu-\mu_{\rm crit}}{\Gamma_{-}})\right],&  \mu_H\leq \mu\leq \mu_{\rm crit}  \\
\end{array} 
\right.
\]
\[
f_{<}(\mu) =  \left\{
\begin{array}{lr}
\frac{1}{2}\left[1-\sin(\frac{\pi}{2}   \frac{\mu-\mu_{\rm crit}}{\Gamma_{+}})\right], &  \mu_{\rm crit}\leq \mu\leq \mu_Q \\
0, &  \mu>\mu_Q \\
\end{array} 
\right. 
\]
where $\mu_H=\mu_{\rm crit}-\Gamma_{-}$ and $\mu_Q=\mu_{\rm crit}+\Gamma_{+}$, see Fig.~\ref{Switches}.
The pressure contribution is modeled also by a trigonometric ansatz,
\[ \Delta(\mu) =  \left\{
\begin{array}{lr}
      0 & : \mu<\mu_H\\
   \frac{1}{2}\Delta P\left[1+\cos(\pi   \frac{\mu-\mu_{\rm crit}}{\Gamma_{-}})\right]& : \mu_H\leq \mu\leq \mu_{\rm crit} \\
   \frac{1}{2}  \Delta P\left[1+\cos(\pi  \frac{\mu-\mu_{\rm crit}}{\Gamma_{+}})\right] & :  \mu_{\rm crit}\leq \mu\leq \mu_Q \\
      0 & :  \mu>\mu_Q \\
\end{array} 
\right. \]
shown in Fig.~\ref{Delta}.
The parameters $\Gamma_{+}$ and $\Gamma_{-}$ are fixed from the conditions that
\begin{eqnarray}
	\label{boundH}
     \frac{\partial^{2}P}{\partial\mu^{2}}\Bigr|_{\mu=\mu_H} &=&   \frac{\partial^{2}P_{H}}{\partial\mu^{2}}\Bigr|_{\mu=\mu_H}~, \\
      \frac{\partial^{2}P}{\partial\mu^{2}}\Bigr|_{\mu=\mu_Q}  &=&   \frac{\partial^{2}P_{Q}}{\partial\mu^{2}}\Bigr|_{\mu=\mu_Q }~,
      \label{boundQ}
\end{eqnarray} 
which are equivalent to\footnote{We note that the MIM can also be applied in the case when a standard Maxwell construction would not make sense since the quark matter pressure would dominate the hadronic pressure at low densities and vice-versa at high densities, so that a transition from quark to hadronic matter would occur when increasing the density, as depicted in Fig.~4 of Ref.~\cite{Baym:2017whm}. 
In this case the MIM would be applied with a negative $\Delta P$ parameter, as follows from Eq.~(\ref{eq:DeltaP}). 
In this case there is a minimal $|\Delta P|$ for which $\partial P/\partial n >0 $,
so that the limit $\Delta P\to 0$ cannot be taken (which would correspond to a Maxwell construction).} 
\begin{eqnarray}
\label{eq:DeltaP}
\Delta P &=& (P_H-P_Q)/4 \bigg|_{\mu=\mu_H}=(P_Q-P_H)/4\bigg|_{\mu=\mu_Q}~,
\end{eqnarray}
in dependence on the only free parameter is $\Delta P$.
An important feature is that this interpolation function extends only over a finite range of chemical potentials, since it switches on at $\mu=\mu_H=\mu_{\rm crit}-\Gamma_-$ where it connects the mixed phase to the purely hadronic EoS and switches off at $\mu=\mu_Q=\mu_{\rm crit}+\Gamma_+$ connecting the mixed phase to the pure quark matter EoS.
This property of the interpolation is a qualitative difference to the interpolation suggested in \cite{Masuda:2012ed}
where the extension of the switch functions is not strictly bounded to a finite range since they are taken as hyperbolic tangens functions.

The necessary basis for the application of the MIM are at least two equations of state, $P_H(\mu)$ and $P_Q(\mu)$, for two phases which are defined in respective domains of chemical potentials with a sufficiently broad overlap. 
This region has to include the crossing of both functions which defines the critical chemical potential $\mu_{\rm crit}$ and the critical pressure $P_{\rm crit}=P_H(\mu_{\rm crit})=P_Q(\mu_{\rm crit})$ of the Maxwell construction between both phases.

The consideration of structures in the mixed phase of a first order phase transition as obtained by such a Maxwell construction will be of particular relevance when the jump in the (energy) density at this transition is large. In this situation one has to expect pronounced inhomogeneities and a decisive role of a finite surface tension in the mixed phase.
For a detailed discussion of pasta phases at the hadron-quark phase transition see, e.g., Refs.~\cite{Voskresensky:2002hu,Maruyama:2007ss,Watanabe:2003xu,Horowitz:2005zb,Horowitz:2014xca,Newton:2009zz,Yasutake:2014oxa} for various descriptions.
Such a situation is known to prevail for EoS of hybrid compact star matter that entail the phenomenon of high mass twin (HMT) compact stars \cite{Benic:2014jia} being related to the appearance of a third family branch of hybrid stars in the mass-radius diagram, disconnected from the second family branch of ordinary neutron stars.
In this case there exists a range of star masses for which pairs of compact stars have the same mass but different radii. 
Typically, pure hadronic compact stars will populate the second branch while hybrid compact stars will be located on the third family, whereas white dwarfs are located on the first branch. 
The HMT case is of great importance because it bears the possibility of probing the existence of a critical endpoint in the QCD phase diagram~\cite{Alvarez-Castillo:2015xfa}, serving as a guide for estimating heavy ion collision conditions~\cite{Alvarez-Castillo:2016wqj} as well as solving different issues on the compact star phenomenology~\cite{Blaschke:2015uva}. 

\section{A hybrid compact star EoS using the MIM}

In this work we base our mixed phase calculations on an HMT EoS in the form of a piecewise polytropic representation \cite{Hebeler:2013nza,Alvarez-Castillo:2017qki,Read:2008iy,Raithel:2016bux} of the neutron star matter EoS at supersaturation densities ($n_1<n<n_5\gg n_0$)
%---------------------------------------------------------
\begin{eqnarray}
\label{polytrope}
P(n) =  \kappa_i  (n/n_0)^{\Gamma_i}, \,  n_i < n < n_{i+1}, \,  i=1 \dots 4~.
\end{eqnarray} 
%---------------------------------------------------------
Here $\Gamma_i$ is the polytropic index of each of the density regions
labeled by $i=1\dots 4$.  The first polytrope correspond to a stiff
nucleonic EoS. The second polytrope represents a first order phase
transition with a constant pressure $P=P_{\rm crit}=\kappa_2$ because $\Gamma_2=0$. 
The polytropes in the density regions 3 and 4 above the phase
transition correspond to stiff quark matter. 
The EoS parameters corresponding to the 4-polytrope EoS labelled "ACB4" in 
Ref.~\cite{Paschalidis:2017qmb} are shown in table~\ref{tab:1}.

\begin{table}[!thb]
	\centering
	\caption{%\color{red} 
		Parameters for the 4-polytrope hybrid EoS model of Eq.~\ref{polytrope} corresponding to set 4 of 
		Ref.~\cite{Alvarez-Castillo:2017qki} labelled "ACB4" in Ref.~\cite{Paschalidis:2017qmb}.  }
%	\label{param-123}
	\begin{tabular}{c|cccc|c}
		\hline \hline
		&$\Gamma_i$&$\kappa_i$&$n_i$ &$m_{0,i}$&$M_{\rm max/min}$\\		
		i&&[MeV/fm$^3$]&[1/fm$^3$] &[MeV]&[M$_\odot$]\\		
\hline
		1&4.921&2.1680&0.1650&939.56 & 2.01  \\
		2&0.0&63.178&0.3174&939.56 & -- \\
		3&4.000&0.5075&0.5344&1031.2 & 1.96  \\
		4&2.800&3.2401&0.7500&958.55 & 2.11  \\
		\hline \hline
	\end{tabular}
	\label{tab:1}
\end{table}

For the present interpolation construction we need to convert the EoS (\ref{polytrope}) 
to the form \cite{Alvarez-Castillo:2017qki}
\begin{equation}
\label{P-mu}
P(\mu)=\kappa_i\left[(\mu - m_{0,i})\frac{\Gamma_i -1}{\kappa_i\Gamma_i} \right]^{\Gamma_i/(\Gamma_i-1)} ~,
\end{equation}
valid for the respective regions (phases) $i=1\dots 4$, where for the constant pressure region $i=2$
this formula collapses to $P(\mu=\mu_{\rm crit})=P_{\rm crit}=\kappa_2$ because of $\Gamma_2=0$.
For applying the MIM (\ref{Interpolation}) it is important that the pressure of the hadronic phase ($i=1$) 
valid for $\mu<\mu_{\rm crit}$ can be extrapolated to the neighboring quark matter phase ($i=3$) where 
 $\mu>\mu_{\rm crit}$ and vice-versa.

\begin{figure}[!th]
	\includegraphics[width=0.5\textwidth]{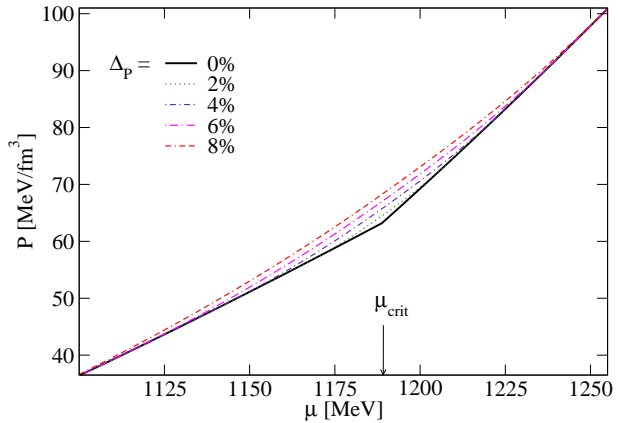}
	%\vspace{1cm}
\caption{The pressure $P$ as a function of chemical potential $\mu$ where the MIM is employed. According to the legend each line corresponds to a value of the dimensionless increment of the pressure $\Delta_P=\Delta P/P_{\rm crit}=0\, (0.02)\, 0.08$ . The Maxwell construction is obtained for $\Delta_P=0$.} 
\label{fig:PvsMu} 
\end{figure}

In Fig.~\ref{fig:PvsMu} we show the hybrid EoS $P(\mu)$ resulting from employing the MIM. 
As indicated in the legend each line corresponds to a value of the dimensionless increment of the pressure $\Delta_P=\Delta P/P_{\rm crit}=0\, (0.02)\, 0.08$ . 
The case $\Delta_P=0$ corresponds to the Maxwell construction where pasta phase effects are absent. 
The critical chemical potential is $\mu_{\rm crit}=1189$ MeV and $P_{\rm crit}=63.18$ MeV/fm$^3$  in this case.
The extension of the mixed phase region defined by the parameters $\Gamma_-$ and $\Gamma_+$ as solutions of Eqs.~(\ref{boundH}) and (\ref{boundQ}) is given for a set of $\Delta_P$ values in 
table~\ref{tab:2}.

\begin{table}[!hbt]
	\centering
	\caption{%\color{red} 
		 The extension of the mixed phase region defined by the parameters $\Gamma_-$ and $\Gamma_+$ as solutions of Eqs.~(\ref{boundH}) and (\ref{boundQ}) is given for different values of the dimensionless pressure increment
		$\Delta_P$.  }
	\begin{tabular}{ccc}
		\hline \hline
		$\Delta_P$&$\Gamma_-$&$\Gamma_+$\\		
		& [MeV]&[MeV]\\		
		\hline
		0.01&	11.9126&	11.2956\\
		0.02&	24.4359&   22.0175\\
		0.03&	37.7194&   32.2609\\
		0.04&	51.9709&   42.0965\\
		0.05&	67.4957&   51.5787\\
		0.06&	84.8171&	60.7503\\
		0.07&	104.910&	69.6458\\
		0.08&	130.078&	78.2934\\	
		\hline \hline
	\end{tabular}
\label{tab:2}
\end{table}

\begin{figure}[!ht]
	\includegraphics[width=0.5\textwidth]{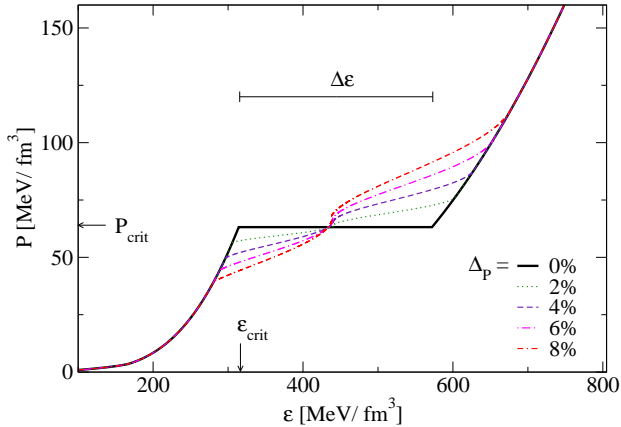}
	%\vspace{1cm}
\caption{The pressure $P$ as a function of energy density $\varepsilon$ for hybrid star matter mimicking pasta phase effects according to the MIM of the present work. Different lines stand for $\Delta_P=0\, (0.02)\, 0.08$, with  $\Delta_P=0$ corresponding to the Maxwell construction to be recognized on the horizontal pressure branch.} 
\label{fig:PvsEpsilon} 
\end{figure}

\begin{figure}[!htb]
	\includegraphics[width=0.5\textwidth]{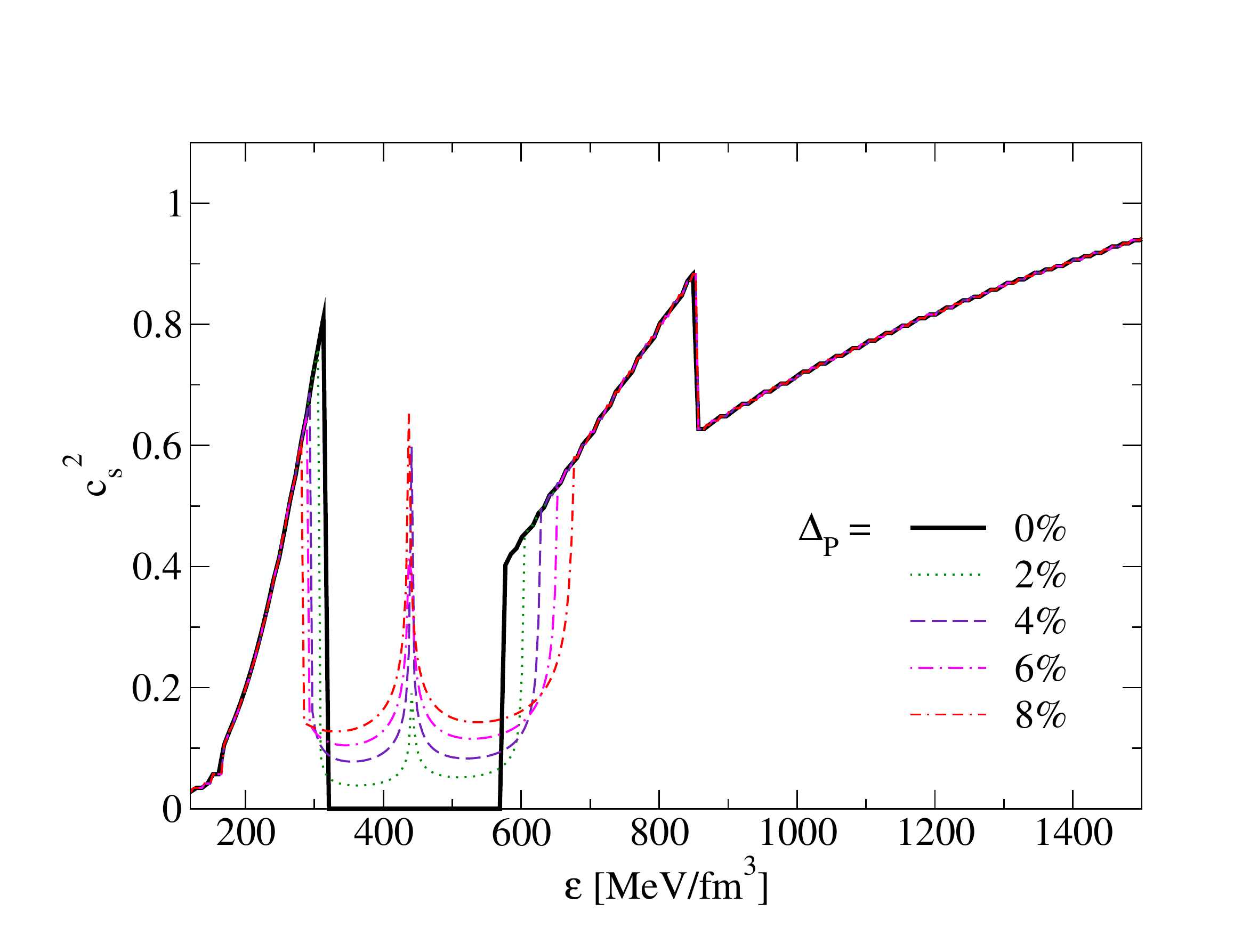}\\
	%\vspace{1cm}
\caption{Squared speed of sound as a function of the energy density for the MIM. It features an asymmetric peak at the middle of the mixed phase region as a characteristic feature of the MIM.
Different lines stand for $\Delta_P=0\, (0.02)\, 0.08$, with  $\Delta_P=0$ corresponding to the Maxwell construction
that entails vanishing speed of sound in the mixed phase.} 
\label{fig:SoS} 
\end{figure}

In Fig.~\ref{fig:PvsEpsilon} we show the EoS of the hybrid star matter in the form $P(\varepsilon)$ which exhibits the effect of the MIM in mimicking pasta phase effects, for different values of the single parameter $\Delta_P$, the dimensionless pressure increment. 
Note that in comparison to a simple interpolation that replaces a region around the critical pressure of the Maxwell construction in $P(\mu)$ by a parabolic (polynomial) function \cite{Ayriyan:2017nby}, we observe an intermediate stiffening of the EoS. 
This reminds of the behavior of the substitutional compounds in the neutron star crust, see Fig.~\ref{Compound}. 
This intermediate stiffening is most clearly demonstrated when considering the squared speed of sound 
%$c_s^2=\frac{\partial P}{\partial \varepsilon}$ 
as a function of the energy density in Fig.~\ref{fig:SoS} which shows a peak in the middle of the phase transition region that is the more pronounced the larger the value of $\Delta_P$ is.

\begin{figure}[!htb]
\includegraphics[width=0.5\textwidth]{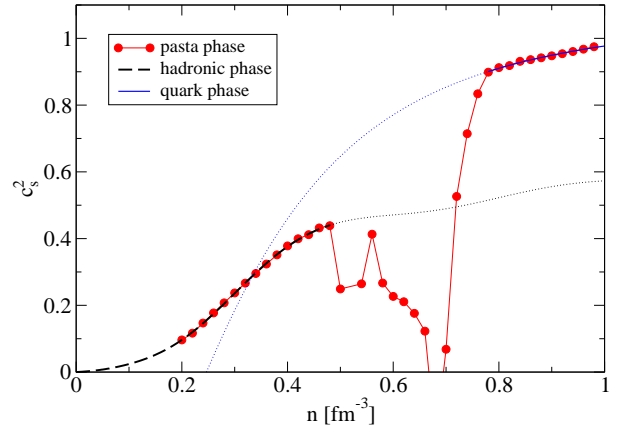}
\caption{Realistic mixed phase calculation including geometrical structures based on~\cite{Maslov:2018ghi}. The phase transition between the hadron and the quark phases features a non-monotonous speed of sound $c_{s}^{2}$ structure of the substitutional compound type which is properly described by the mixing interpolation method (MIM) presented in this work.} 
\label{cs2_s40_nomuons} 
\end{figure}

The question arises whether such an intermediate stiffening can be a real effect that is due to the account for finite-size structures with surface tension $\sigma$ in the mixed phase.
We answer this question affirmative and refer to the behaviour of $c_s^2(\varepsilon)$ for the example of a pasta phase construction that has been performed in Ref.~\cite{Maslov:2018ghi} between a hadronic EoS (KVORcut03) and a quark matter EoS (SFM, $\alpha=0.3$) for a surface tension $\sigma = 40$ MeV/fm$^2$
(corresponding to the case H2-Q1 in Fig.~3 of \cite{Maslov:2018ghi}), for which we derived the squared speed of sound and show it in Fig.~\ref{cs2_s40_nomuons}. 
We note the qualitative agreement between this figure and Fig.~\ref{fig:SoS} of the MIM presented in this work which both show a peak structure in the middle of the mixed phase of the hadron to quark matter transition.
We attribute this intermediate stiffening  to a commensurate structure of nucleons and quark matter droplets that could be pictured like the case of substitutional compounds shown in Fig.~\ref{Compound}, where the open circles correspond to the nucleons and the filled circles to the quark droplets. 

An important condition for the HMTs to appear is the Seidov constraint
\cite{Seidov:1971}
\begin{equation}
\label{seidov}
\frac{\Delta\varepsilon}{\varepsilon_{\rm crit}} \ge \frac{1}{2} + \frac{3}{2} \frac{P_{\rm crit} }{\varepsilon_{\rm crit}} ~,
\end{equation}
which corresponds to a straight line in the so-called "phase diagram of hybrid stars"
\cite{Alford:2013aca}.
In the upper half-plane of that diagram where (\ref{seidov}) is fulfilled, the onset of the phase transition in the compact star center leads to a gravitational instability of the configuration. 
Eventually, for a sufficiently stiff high-density part of the EoS, the hybrid star will resume stability until a maximum mass is reached which shall exceed the present observational constraint of $2.01$ M$_\odot$ \cite{Antoniadis:2013pzd}. 
While being stiff enough to fulfill the maximum mass constraint, the EoS must fulfill the causality constraint $c_s<c$ for the speed of sound $c_s=\sqrt{\partial P/\partial \varepsilon}$ at all densities attained in the stable star configurations.

%%%%%%%%%%%%%%%%Figures%%%%%%%%%%%%%%%%

In the following we would like to apply the MIM to the case of the HMT stars and test the robustness of the HMT phenomenon against a broadening of the mixed phase region by the MIM.

\section{Sequences of hybrid star configurations}

Once we have derived the mixed phase EoS we solve the Tolman--Oppenheimer--Volkoff (TOV) equations 
~\cite{Tolman:1939jz,Oppenheimer:1939ne}
\begin{eqnarray}
\frac{dP( r)}{dr}&=& -\frac{G\left(\varepsilon( r)+P( r)\right) \left(M( r)+ 4\pi r^3 P( r)\right)}{r\left(r- 2GM( r)\right)},\\
\frac{dM( r)}{dr}&=& 4\pi r^2 \varepsilon( r),
\end{eqnarray}
that describe static, spherically-symmetric compact stars in the framework of Einstein's general relativity theory.

\begin{figure*}[!h]
	\begin{center}$
		\begin{array}{cc}
		\includegraphics[height=0.45\textwidth]{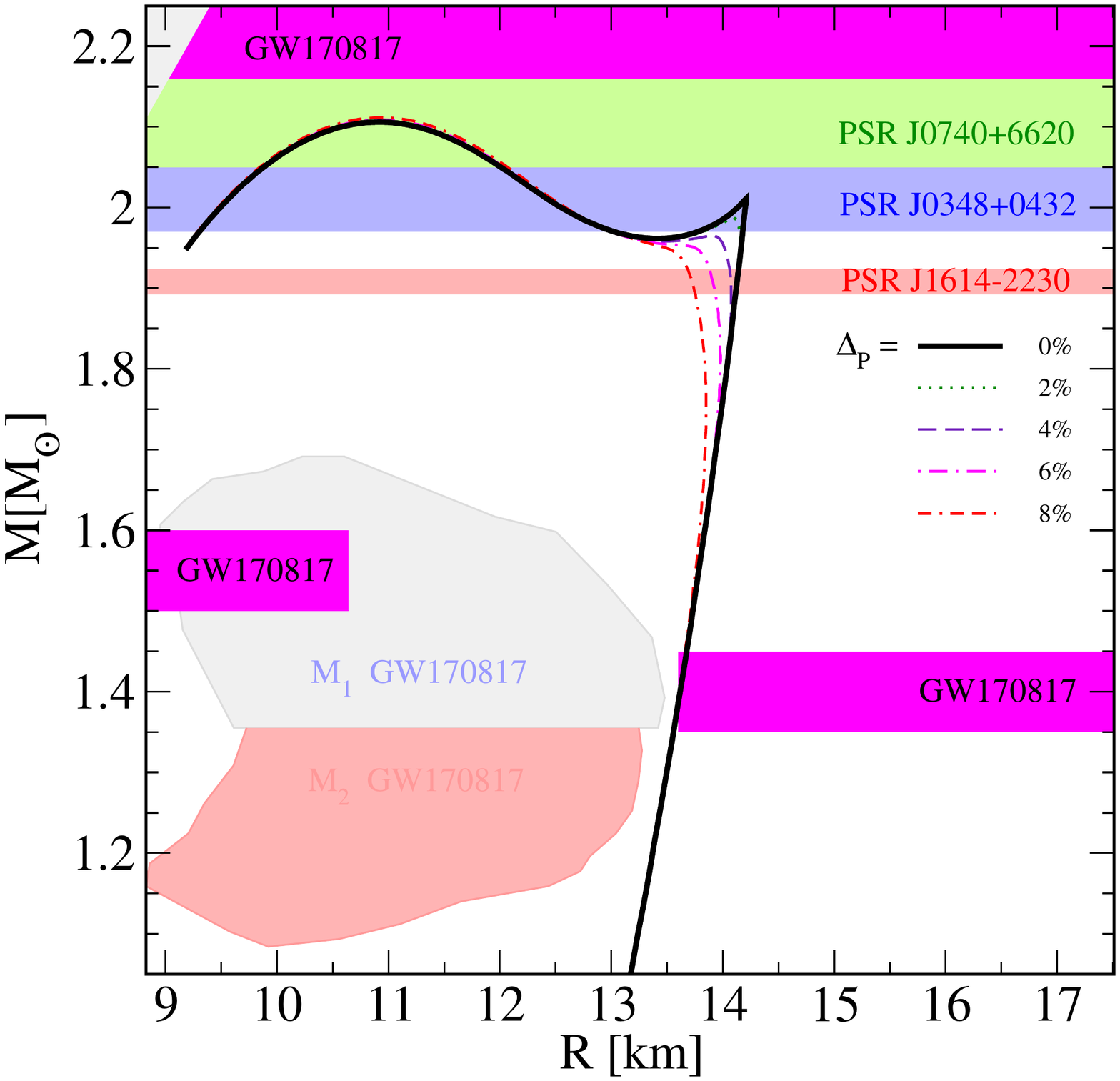}
&\hspace{-1cm}	
		\includegraphics[height=0.45\textwidth]{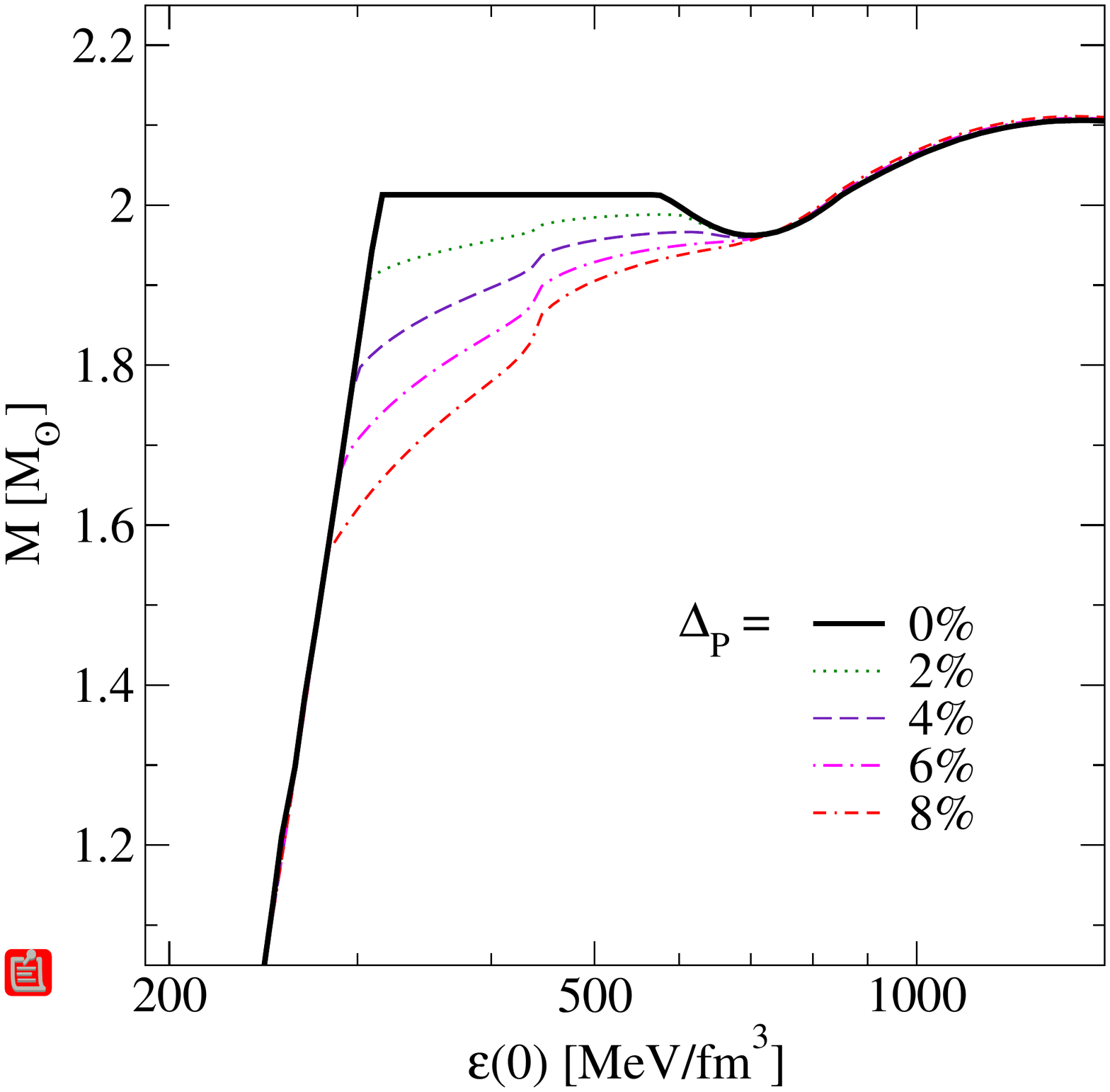} 	
				%\vspace{1cm}
		\end{array}$
	\end{center}
	\vspace{-0.5cm}
	\caption{\label{MvsR}
	Left panel: Mass-radius relation exhibiting the HMT phenomenon depending on the pressure increment 
	$\Delta_P=0\, (0.02)\, 0.08$ of the MIM. 
	For $\Delta_P\ge 6$\% the HMT phenomenon is lost and the hybrid star branch gets connected with the neutron star branch. The upper colored bands represent the mass measurements for PSR J1614-2230~\cite{Demorest:2010bx,Fonseca:2016tux,Arzoumanian:2017puf}, PSR J0348+0432~\cite{Antoniadis:2013pzd},  and PSR J0740+6620 \cite{Cromartie:2019kug} as the most massive compact stars. 
%		The green region represents the mass of the nearest millisecond pulsar PSR J0437-4715  \cite{Reardon:2015kba} for which a radius measurement is under way by the NICER experiment \cite{NICER:2014}. 
	The regions for $\textrm{M}_1$ and $\textrm{M}_2$ were obtained by a low-spin prior analysis of 
		%the gravitational wave signal from the compact star merger event 
	GW170817 \cite{TheLIGOScientific:2017qsa}. 
	The three hatched regions labeled "GW170817" are excluded by the merger event and constrain the minimal radius at $M=1.6$ M$_\odot$~\cite{Bauswein:2017vtn}, the maximal radius at $M=1.4$ M$_\odot$~\cite{Annala:2017llu} and the upper limit on the maximum mass~\cite{Rezzolla:2017aly}.
	Right panel: Effects of the mixed phase on the gravitational mass $M$ is as a function of the central energy density 
	$\varepsilon(0)$ obtained with the MIM for different values of the dimensionless pressure increment 
	$\Delta_P=0\, (0.02)\, 0.08$. 
	} 
\end{figure*}

The solution of this coupled system of differential equations
proceeds with dialling a value for the central pressure of the
star configuration,  $ P_c= P(r=0)$, and then integrating outwards
until the pressure drops to zero, $P(r=R)=0$, which defines the
radius $R$ of the star. 

Simultaneos integration of the second equation
yields the gravitational mass of the star enclosed in the spherical 
matter distribution $M(r)$ up to the actual distance $r$ from the center.
Reaching the radius of the star the integration is stopped and the
enclosed mass equals the total gravitational mass $M=M(r=R)$
of the star.
The solutions provide internal pressure profiles $P(r)$ of the 
stars which with the knowledge of the EoS $P(\varepsilon)$ can be 
converted to profiles of energy density. 
Once with the central pressure $P_c=P(\varepsilon_c)$ also the 
central density $\epsilon_c$ is chosen, the triple of numbers
$(\varepsilon_c,M,R)$ can be plotted pairwise against each other 
thus characterising uniquely the EoS. 
The main result of such calculations is the mass-radius $(M-R)$
diagram shown on the left panel of Fig.~\ref{MvsR} for the set of hybrid EoS obtained 
with the MIM when varying the free parameter $\Delta_P$ of this 
mixed phase construction.
We find that the HMT phenomenon of the input EoS is rather robust against
mixed phase effects here mimicked by applying the MIM. 
Only for $\Delta_P \ge 6$ \% the twin phenomenon vanishes since the third 
family sequence representing hybrid stars joins the second one of neutron stars.
This is in accordance with earlier investigations of the HMT 
robustness within the RIM \cite{Ayriyan:2017nby}.
The obtained $M-R$ sequences shown in Fig.~\ref{MvsR} are all in accordance with the 
constraints known so far.
This includes the most recent constraints derived from GW170817
that are shown as exclusion regions labeled "GW170817" in Fig.~\ref{MvsR}. 
They concern the minimal radius at $M=1.6$ M$_\odot$~\cite{Bauswein:2017vtn}, 
the maximal radius at $M=1.4$ M$_\odot$~\cite{Annala:2017llu} and the 
upper limit on the maximum mass~\cite{Rezzolla:2017aly} of TOV sequences.

\begin{figure}[!h]
\includegraphics[width=0.5\textwidth]{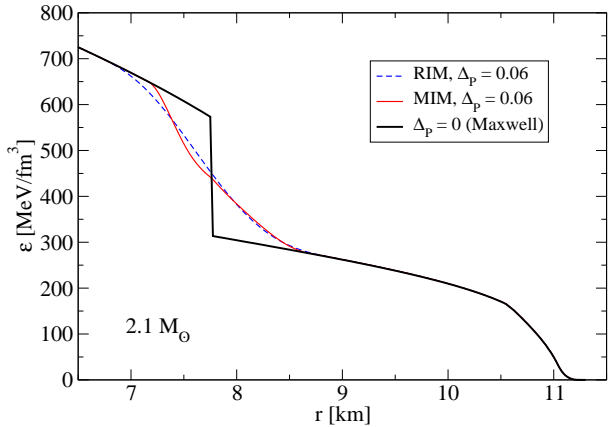}
	\caption{\label{MRvsnc} 
	Comparison of energy density profiles for a 2.1 M$_\odot$ hybrid compact star for three cases: both methods of defining the mixed phase (RIM and MIM) for a pressure increment $\Delta_P=0.06$ and the limiting case of $\Delta_P=0$ (Maxwell construction) without a mixed phase. Note the suppressed zero of the $r-$ axis.
	} 
\end{figure}

On the mass-radius diagram of Fig.~\ref{MvsR}, however, the difference between the RIM and the MIM 
treatment of the mixed phase construction is hardly recognizeable as the direct comparison of both methods 
in Ref.~\cite{Abgaryan:2018gqp} shows.
A difference between both approaches could manifest itself in more subtle observable effects, such as the
cooling behavior of hybrid stars. 
It has been discussed, e.g., in Ref.~\cite{Grigorian:2016leu,Grigorian:2017owg} the density profile of pairing gaps has a 
strong influence on the cooling history of neutron stars in the temperature-age diagram. 
Consequently, the modification of the density profile of the star in a sensible region shall manifest itself in 
a noticeable change of the expected cooling history. 
In the right panel of Fig.~\ref{MvsR} we show the mass as a function of the central energy density for different
values of the mixed-phase parameter $\Delta_P=0\, (0.2)\, 0.8$ using the MIM. 
We observe that for a pronounced intermediate stiffening effect (larger $\Delta_P$ values) there is a step-like
increase in the mass for the central energy density region in the middle of the mixed phase, where the 
speed of sound peaks. 
Coming to the direct comparison with the RIM  \cite{Abgaryan:2018gqp,Ayriyan:2017nby}, 
we show in Fig.~\ref{MRvsnc} the energy density profile of a hybrid star with $M=2.1~M_\odot$
for $\Delta_P = 0.06$ using the RIM and MIM, respectively. 
The RIM uses polynomial interpolation functions, here a parabolic interpolation between hadronic and quark matter pressures in the $P(\mu)$ plane is chosen.
When contrasting the (energy)density profiles of the RIM and MIM with the one obtained by a Maxwell construction
(see the black solid line in Fig.~\ref{MRvsnc}), the strongest effect is obtained due to the step of accounting for the mixed phase at all.
The difference between using the RIM or MIM to mimic pasta phases is a secondary effect. 
Nevertheless it would be of interest to use the solutions for the star structure obtained here and to perform a systematic investigation of the cooling behavior in order to estimate quantitatively to what extent the difference in the treatment of the quark-hadron mixed phase affects the hybrid star cooling.  

%%%%TOV %%%%%

\section{Conclusion}
In this work we have developed a new interpolation method to mimic pasta phase effects in the course of a strong first-order phase transition. 
This mixing interpolation method superimposes the equations of state of the subphases with weight functions 
in a bounded region around the critical chemical potential of the corresponding Maxwell construction. 
This method results in an intermediate stiffening of the equation of state in the middle of the mixed phase region which may be attributed to a "locking" of commensurate subphase structures similar to the situation in substitutional compound structures in the neutron star crust.
The application of this method to the case of a high-mass twin star equation of state confirms the earlier finding within the replacement interpolation method \cite{Ayriyan:2017tvl} that the twin phenomenon is rather robust against mixed phase effects.
We conclude that this new mixing interpolation method may provide a tool to fit and interpret full pasta phase calculations in some cases more appropriately than the replacement interpolation method for which the intermediate stiffening effect is absent. 
An observable effect of this difference between both approaches could manifest itself in the cooling behavior of hybrid 
compact stars. 

\subsection*{Acknowledgements}
D.B. is grateful for the hospitality and spiritual atmosphere of the Orthodox Academy of Crete in Kolymbari where the main part of this work has been completed. He acknowledges T. Tatsumi, D.N. Voskresensky and N. Yasutake for discussions on the pasta phase construction, and A. Ayriyan, N. Chamel, H. Grigorian and K. Maslov for valuable comments.  
The authors thank K. Maslov and N. Yasutake for providing them with the data for figure \ref{cs2_s40_nomuons}. 
The work of D.B. was supported in part by the Russian Science Foundation under contract number 17-12-01427. 
D.E.A-C. is grateful to the Bogoliubov-Infeld program for supporting the collaboration and scientist exchange between JINR Dubna and Polish Institutes. 
%

 %%%%%%%%%%%%%%%%%%%%%

\end{document}